\documentclass[preprint,prd,nofootinbib,tightenlines,amsmath]{revtex4}

\usepackage{axodraw} 
\usepackage{bm} 
\usepackage{graphicx}
    
\begin{document}

\baselineskip=15pt
\parskip=5pt
        
\hfill SMU-HEP-03-11
     
 
\vspace*{10ex}
      
\title{New Physics and \mbox{$\bm{CP}$}  Violation in Hyperon 
Nonleptonic Decays}  
     
\author{Jusak Tandean}
\email{jtandean@mail.physics.smu.edu}

\affiliation{  
Department of Physics, Southern Methodist University,   
Dallas, Texas 75275-0175 
\vspace{5ex} \\}
    

\begin{abstract}  
The sum of the $CP$-violating asymmetries $A(\Lambda^0_-)$  and  $A(\Xi^-_-)$  
in hyperon nonleptonic decays is presently being measured by the E871 experiment.   
We evaluate contributions to the asymmetries induced by chromomagnetic-penguin 
operators, whose coefficients can be enhanced in certain models of new physics.  
Incorporating recent information on the strong phases in  $\Xi\to\Lambda\pi$  
decay, we show that new-physics contributions to the two asymmetries can 
be comparable. 
We explore how the upcoming results of E871 may constrain the coefficients 
of the operators. 
We find that its preliminary measurement is already better than the $\epsilon$ 
parameter of $K$-$\bar{K}$  mixing in bounding the parity-conserving contributions.   
\end{abstract}   
     
\pacs{}
    
\maketitle

\section{Introduction\label{intro}} 
    
The phenomenon of  $CP$  violation remains one of the least understood 
aspects of particle physics.  
Although $CP$ violation has now been detected in several processes 
in the kaon and $B$-meson systems~\cite{cpx}, its origin is still far 
from clear.   
The standard model (SM) can accommodate the experimental results, but 
they do not yet provide critical tests for it~\cite{cpreview}.  
To pin down the sources of $CP$ violation within or beyond the SM, 
it is essential to observe it in many other processes.

Nonleptonic hyperon decays provide an environment where it is possible 
to obtain additional observations of $CP$ violation. 
Although this has been recognized for a long time~\cite{op}, only 
recently has it been experimentally feasible to search for 
$CP$ violation in some of these decays~\cite{hcpx,hypercp,zyla}.  
There is currently such an effort being done at Fermilab, 
by the HyperCP~(E871) Collaboration~\cite{hypercp,zyla}.

The reactions studied by HyperCP are the decay sequence 
$\,\Xi^-\to\Lambda\pi^-,\,$  $\,\Lambda\to p\pi^-\,$  and 
its antiparticle counterpart.  
For each of these decays, the decay distribution in the rest frame of 
the parent hyperon with known polarization  $\bm{w}$ has the form   
\begin{equation}   
\frac{{\rm d}\Gamma}{{\rm d}\Omega}  \,\sim\,  
1 +\alpha\, \bm{w}\cdot\hat{\bm{p}}   \,\,,  
\end{equation}   
where  $\hat{\bm{p}}$  is the unit vector of the daughter-baryon momentum 
and  $\alpha$  is the parameter relevant to the $CP$ violation 
of interest.   
By evaluating the decay chain  $\,\Xi\to\Lambda\pi\to p\pi\pi,\,$
the HyperCP experiment is sensitive to the {\sl sum} of $CP$ violation in 
the  $\Xi$  decay and  $CP$ violation in the  $\Lambda$  decay.  
Thus it measures~\cite{hypercp,zyla} 
\begin{equation}
A_{\Xi\Lambda}^{}  \,=\,  A_\Lambda^{} + A_\Xi^{}  \,\,,
\label{adefin}
\end{equation} 
where   
\begin{equation}
A_\Xi^{}  \,\equiv\,  A\bigl(\Xi^-_-\bigr)  \,\equiv\,  
\frac{\alpha_\Xi^{}+\overline{\alpha}_\Xi^{}}{
\alpha_\Xi^{}-\overline{\alpha}_\Xi^{} }   \,\,,   
\hspace{3em}  
A_\Lambda^{}  \,\equiv\,   
A\bigl(\Lambda^0_-\bigr)  \,\equiv\,  
\frac{\alpha_{\Lambda}^{}+\overline{\alpha}_{\Lambda}^{}}{
\alpha_{\Lambda}^{}-\overline{\alpha}_{\Lambda}^{} }   \,\,,   
\end{equation}  
are the $CP$-violating asymmetries in  $\,\Xi\to\Lambda\pi\,$   
and  $\,\Lambda\to p\pi,\,$  respectively.  
The published measurements currently available~\cite{pdb} are 
$\,A_\Lambda^{}=0.012\pm0.021\,$  and  $\,A_{\Xi\Lambda}^{}=0.012\pm 0.014.\,$  
HyperCP will obtain more precise results, with an expected sensitivity  
of~$\sim$$10^{-4}$,  and has recently reported~\cite{zyla} a preliminary  
measurement of  $\,A_{\Xi\Lambda}^{}=(-7\pm12\pm6.2)\times 10^{-4}.\,$

The amplitudes for both  $\,\Lambda\to p\pi^-\,$ and  $\,\Xi^-\to\Lambda\pi^-\,$  
contain  $S$- and $P$-wave components, each of which consists of contributions 
describing  $\,|\Delta I|=\frac{1}{2}$  and  $\frac{3}{2}\,$  
transitions, with the former being known to dominate. 
Assuming $\,|\Delta I|=\frac{1}{2}\,$  dominance, one can derive at 
leading order~\cite{dhp}   
\begin{equation}   \label{approxa}
\begin{array}{c}   \displaystyle
A_{\Lambda}^{}  \,=\,    
- \tan \left(\delta_P^\Lambda -\delta_S^\Lambda\right) 
\sin\left(\phi_P^\Lambda-\phi_S^\Lambda\right)  \,\,,  
\hspace{3em} 
A_{\Xi}^{}  \,=\,    
- \tan \left(\delta_P^\Xi -\delta_S^\Xi\right) 
\sin\left(\phi_P^\Xi-\phi_S^\Xi\right)  \,\,.  
\end{array}
\end{equation}
Here, $\delta_{S}^\Lambda$ $\bigl(\delta_{P}^\Lambda\bigr)$  is the 
strong $S$-wave ($P$-wave)  $N\pi$ scattering phase-shift at  
$\,\sqrt s=M_\Lambda^{},\,$  and  
$\delta_{S}^\Xi$ $\bigl(\delta_{P}^\Xi\bigr)$ 
is the strong $S$-wave ($P$-wave) $\Lambda \pi$ scattering 
phase-shift at $\,\sqrt s=M_\Xi^{}.\,$ 
Moreover,  $\phi_{S}^{\Lambda,\Xi}$  $\bigl(\phi_{P}^{\Lambda,\Xi}\bigr)$  
are the $CP$-violating weak phases induced by the $\,|\Delta S|=1,\,$  
$|\Delta I|=\frac{1}{2}\,$  interaction in the $S$-wave ($P$-wave) of 
the  $\,\Lambda\to p\pi^-\,$ and  $\,\Xi^-\to\Lambda\pi^-\,$  decays, 
respectively.

The strong  $N\pi$  scattering phases needed in Eq.~(\ref{approxa}) 
have been measured~\cite{roper} to be  $\,\delta_S^\Lambda\sim 6^\circ\,$  
and  $\,\delta_P^\Lambda\sim -1^\circ\,$  with errors of 
about~$1^\circ.\,$ 
In contrast, the strong  $\Lambda\pi$ phases are less well 
determined.  
Using the current PDG numbers~\cite{pdb}, one can deduce~\cite{huang} 
the phase difference  $\,\delta_P^\Xi-\delta_S^\Xi=(-7.7\pm7.7)^\circ.\,$  
Very recently, the E756 Collaboration~\cite{e756'} has published  
a new measurement of  
$\,\delta_P^\Xi-\delta_S^\Xi=(+3.17\pm5.28\pm0.73)^\circ.\,$  
HyperCP is presently also measuring this quantity, with better 
precision, and has reported~\cite{huang} a preliminary result of      
$\,\delta_P^\Xi-\delta_S^\Xi=\bigl(7.6\pm1.3_{-2.8}^{+2.4}\bigr)^\circ.\,$

On the theoretical side, the most recent update~\cite{tv} of 
the standard-model prediction of~$A_{\Xi\Lambda}^{}$  yields a value  
that is smaller than most of earlier estimates~\cite{dhp,hsv}, 
but with a sizable uncertainty, resulting in the range  
$\,\bigl|A_{\Xi\Lambda}^{}\bigr|\lesssim 5\times10^{-5},\,$  
which is compatible with some of the earlier predictions.
Thus, the upcoming data from HyperCP will likely be insensitive to 
SM effects.       
Beyond the SM, the asymmetry is potentially more detectable, as various 
estimates~\cite{NP,NP'} indicate that  $A_\Lambda^{}$  could be as large 
as $10^{-3}$  in models with enhanced chromomagnetic-penguin operators (CMO).  
In these studies, the corresponding value of  $A_\Xi^{}$  has 
been neglected because most of recent calculations based on chiral 
perturbation theory~\cite{deltax,ttv} suggest that  
$\delta_P^\Xi-\delta_S^\Xi$  is  small compared to  
$\delta_P^\Lambda-\delta_S^\Lambda.\,$     
However, there are early indications, from a coupled-channel $K$-matrix 
estimate given in Ref.~\cite{ttv} and from the preliminary result of HyperCP 
above~\cite{huang}, that the two phase-differences may actually be comparable 
in size.  
This is also compatible with the other two measurements of  
$\delta_P^\Xi-\delta_S^\Xi$  mentioned earlier.  
Since HyperCP is sensitive only to the sum  $\,A_\Lambda^{}+A_\Xi^{},\,$  
it is therefore important to have an up-to-date expectation of  
$A_{\Lambda,\Xi}^{}$  from possible new physics and of their sum.

In this paper, we estimate both  $A_\Lambda^{}$  and  $A_\Xi^{}$  due 
to possible physics beyond the SM, incorporating the new information on 
the strong phases and taking into account constraints from kaon data.    
Specifically, we consider contributions generated by the CMO, which could be 
significantly larger that their SM counterparts~\cite{NP,NP'}.   
The relevant effective Hamiltonian can be written as~\cite{buras}
\begin{eqnarray}   \label{H_sdg}
{\cal H}_{\rm w}^{}  \,=\,  
C_{g}^{}\, Q_g^{}  \,+\,  \tilde{C}_{g}^{}\, \tilde Q_g^{}   
\,\,+\,\,  {\rm H.c.}   \,\,,    
\end{eqnarray}      
where $C_{g}^{}$  and  $\tilde{C}_{g}^{}$  are the Wilson coefficients, 
and   
\begin{eqnarray}   
Q_{g}^{}  \,=\,  
\frac{g_{\rm s}^{}}{16\pi^2}\, \bar d\, \sigma^{\mu\nu} t^a\, 
\bigl(1+\gamma_5^{}\bigr) s\, G_{\!\mu\nu}^{a}   \,\,,   
\hspace{3em}  
\tilde Q_{g}^{}  \,=\,  
\frac{g_{\rm s}^{}}{16\pi^2}\, \bar d\, \sigma^{\mu\nu} t^a\, 
\bigl(1-\gamma_5^{}\bigr) s\, G_{\!\mu\nu}^{a}  \,\,,  
\end{eqnarray}      
are the CMO, with  $G_a^{\!\mu\nu}$  
being the gluon field-strength tensor  and   
$\,{\rm Tr}\bigl(t^a t^b\bigr)=\frac{1}{2}\delta^{ab}.\,$   
These operators also contribute to the $CP$-violating parameters  
$\epsilon$ in kaon mixing and $\epsilon'$ in kaon decay, as well as to 
other hyperon and kaon observables~\cite{sdg,evenparity,buras}.  
Although  $\epsilon$, $\epsilon'$, and  $A_{\Lambda,\Xi}^{}$ receive 
contributions from the same  $\,|\Delta S|=1\,$  interaction, they probe 
different parts of it.  
Whereas  $\epsilon$ and $\epsilon'$ are sensitive only to parity-even 
and parity-odd contributions, respectively, $A_{\Lambda,\Xi}^{}$ are 
sensitive to both.  
Thus, with  $\epsilon$ and $\epsilon'$  now being well measured,  
we will estimate the range of  $A_{\Xi\Lambda}^{}$  arising from the CMO 
that is allowed by  $\epsilon$ and $\epsilon'$, and then compare it with 
the preliminary result of HyperCP mentioned above.  
Since various new-physics scenarios may contribute differently to 
the coefficients of the operators, we will not focus on specific models,  
but will instead adopt a model-independent approach, only assuming that 
the contributions are potentially sizable.  
Accordingly, we will also explore how well the coefficients can be 
constrained in the event that HyperCP detects no $CP$-violation.

In Sec.~\ref{cpt}, we employ heavy-baryon chiral perturbation theory 
to derive the decay amplitudes at leading order.   
In Sec.~\ref{results}, we calculate the weak phases using matrix elements 
estimated in the MIT bag model. 
We then estimate the $CP$-violating asymmetries, taking into account constraints 
from $CP$ violation in the kaon system, and present a discussion of our results.  
We give our conclusions in Sec.~\ref{conclusion}.

\section{Chiral Lagrangian and decay amplitudes\label{cpt}}
     
The chiral Lagrangian that describes the interactions of 
the lowest-lying mesons and baryons is written down in terms of 
$3\times3$  matrices  $\varphi$  and~$B$  which contain the lightest 
meson-octet and baryon-octet fields, respectively~\cite{bsw,JenMan}.    
The mesons enter through the exponential  
$\,\Sigma=\xi^2=\exp({\rm i}\varphi/f),\,$  where  $f$  
is the pion-decay constant.

In the heavy-baryon formalism~\cite{JenMan}, the baryons in the chiral 
Lagrangian are described by velocity-dependent fields,  $B_v^{}$.   
For the strong interactions, the chiral Lagrangian to lowest 
order in the derivative expansion is given  by~\cite{JenMan}
\begin{eqnarray}   \label{Ls1}   
{\cal L}_{\rm s}^{(1)}  &=&  
\left\langle \bar{B}_v^{}\, {\rm i}v\cdot{\cal D} B_v^{} 
 \right\rangle     
+ 2D \left\langle \bar{B}_v^{} S_v^\mu 
 \left\{ {\cal A}_\mu^{}, B_v^{} \right\} \right\rangle 
+ 2F \left\langle \bar{B}_v^{} S_v^\mu  
 \left[ {\cal A}_\mu^{}, B_v^{} \right] \right\rangle    
\,+\,  
\mbox{$\frac{1}{4}$} f^2 \left\langle  
\partial^\mu\Sigma^\dagger\, \partial_\mu^{}\Sigma \right\rangle    \,\,,  
\end{eqnarray}      
where  $\langle\cdots\rangle$ denotes  ${\rm Tr}(\cdots)$  
in flavor-SU(3) space,  $S_v^{}$  is the spin operator,   
$\,{\cal D}^\mu B_v^{}=\partial^\mu B_v^{}+
\bigl[{\cal V}^\mu,B_v^{}\bigr],\,$     
with  
$\,{\cal V}_\mu^{}=\frac{1}{2} \left( \xi\, \partial_\mu^{}\xi^\dagger 
+ \xi^\dagger\,\partial_\mu^{}\xi  \right) ,\,$   
and   
$\,{\cal A}_\mu^{}=\frac{\rm i}{2}   
\left( \xi\, \partial_\mu^{}\xi^\dagger 
      - \xi^\dagger\, \partial_\mu^{}\xi \right).\,$   
In this Lagrangian, $D$ and $F$  are free parameters which can be 
determined from hyperon semileptonic decays.  
We will adopt the parameter values obtained from fitting tree-level 
formulas~\cite{JenMan}, namely  
$\,D=0.80\,$  and  $\,F=0.50.\,$   
We will also need the chiral Lagrangian that explicitly break chiral 
symmetry~\cite{L2refs}, containing one power of the quark-mass matrix  
$\,M={\rm diag}\bigl(m_u^{},m_d^{},m_s^{}\bigr),\,$   
\begin{eqnarray}   \label{Ls2}   
{\cal L}_{\rm s}^{(2)}  &=&    
\mbox{$\frac{1}{4}$} f^2 \left\langle \chi_+^{} \right\rangle     
\,+\,  
\frac{b_D^{}}{2 B_0^{}} \left\langle \bar B_v^{}   
\left\{ \chi_+^{}, B_v^{} \right\} \right\rangle   
+ \frac{b_F^{}}{2 B_0^{}} \left\langle \bar B_v^{}   
 \left[ \chi_+^{}, B_v^{} \right] \right\rangle     
+ \frac{b_0^{}}{2 B_0^{}} \left\langle \chi_+^{} \right\rangle
 \left\langle \bar B_v^{} B_v^{} \right\rangle    \,\,,   
\end{eqnarray}  
where we have used the notation  
$\,\chi_+^{}=\xi^\dagger\chi\xi^\dagger+\xi\chi^\dagger\xi\,$  
to introduce coupling to external (pseudo)scalar sources,  
$\,\chi= s+ip,\,$ such that in the absence of the external sources  
$\chi$ reduces to the mass matrix,  $\,\chi=2B_0^{} M.\,$   
We will take the isospin limit,  $\,m_u^{}=m_d^{}=\hat m,\,$   and 
consequently  
$\,\chi={\rm diag}\bigl(m_\pi^2,m_\pi^2,2 m_K^2-m_\pi^2\bigr).\,$  
In Eq.~(\ref{Ls2}), the constants  $B_0^{}$, $b_{D,F,0}^{}$ are 
free parameters which can be fixed from data.

In the weak sector, the chiral Lagrangian induced by the chromomagnetic-penguin 
operators has to respect their symmetry properties. 
From  Eq.~(\ref{H_sdg}), we observe that  $Q_g^{}$ and  $\tilde Q_g^{}$  
transform as  $\,\bigl(\bar 3_{\rm L}^{},3_{\rm R}^{}\bigr)\,$  and  
$\,\bigl(3_{\rm L}^{},\bar 3_{\rm R}^{}\bigr),\,$  respectively,  
under  $\,\rm SU(3)_{L}^{}$$\times$$\rm SU(3)_{R}^{}\,$   
transformations.   
Accordingly, the desired Lagrangian at leading order is  
\begin{eqnarray}  
{\cal L}_{\rm w}^{}  &=&     
\beta_D^{} \left\langle \bar B_v^{} 
\left\{ \xi^\dagger h\xi^\dagger, B_v^{} \right\} \right\rangle  
+ \beta_F^{} \left\langle \bar B_v^{} 
\left[ \xi^\dagger h\xi^\dagger, B_v^{} \right] \right\rangle  
+ \beta_0^{} \left\langle h\Sigma^\dagger \right\rangle   
\left\langle \bar B_v^{} B_v^{} \right\rangle 
\,+\,  
\beta_\varphi^{}\, f^2 B_0^{} \left\langle h\Sigma^\dagger \right\rangle   
\nonumber \\ && \!\!\! 
+\,\,  
\tilde\beta_D^{} \left\langle \bar B_v^{} 
\left\{ \xi h\xi,B_v^{} \right\} \right\rangle 
+ \tilde\beta_F^{} \left\langle \bar B_v^{} 
 \left[ \xi h\xi,B_v^{} \right] \right\rangle 
+ \tilde\beta_0^{} \left\langle h\Sigma \right\rangle 
 \left\langle \bar B_v^{} B_v^{} \right\rangle  
\,+\,  
\tilde\beta_\varphi^{}\, f^2 B_0^{} \left\langle h\Sigma \right\rangle   
\,\,+\,\,  {\rm H.c.}   \,\,,
\hspace{2em}  
\label{weakcl} 
\end{eqnarray}  
where  $\beta_i^{}$  $\bigl(\tilde\beta_i^{}\bigr)$  are parameters 
containing the coefficient  $C_g^{}$  $\bigl(\tilde{C}_g^{}\bigr)$, 
and  the  3$\times$3-matrix $h$  selects out  $\,s\to d\,$  
transitions, having elements    
$\,h_{kl}^{}=\delta_{k2}^{}\delta_{3l}^{}.\,$  
As shown in  Appendix~\ref{qbarq}, the expression for ${\cal L}_{\rm w}^{}$  
can be inferred from the lowest-order chiral realization of 
the densities  $\,\bar d(1\pm\gamma_5^{})s.\,$  
We remark that this Lagrangian is of ${\cal O}(1)$ in the derivative 
and $m_s^{}$ expansions.

For the weak decay of a spin-$\frac{1}{2}$  baryon  $B$  into another   
spin-$\frac{1}{2}$  baryon $B'$  and  a pseudoscalar meson~$\phi$, 
the amplitude in the heavy-baryon approach has the general 
form~\cite{jenkins2}    
\begin{eqnarray}   \label{M,HB}     
{\rm i} {\cal M}_{B\to B'\phi}^{}  \,=\,    
-{\rm i} \bigl\langle B'\phi \bigr| {\cal L}_{\rm w+s}^{} 
\bigl| B \bigr\rangle    
\,=\,    
\bar u_{B'}^{}\, \Bigl( {\cal A}_{BB'\phi}^{(S)} 
+ 2S_v^{}\!\cdot\!p_\phi^{}\, {\cal A}_{BB'\phi}^{(P)} 
\Bigr) \, u_{B}^{}   \,\,,  
\end{eqnarray}    
where the superscripts refer to the  $S$- and $P$-wave components 
of the amplitude.  
These components are related to the decay width  $\Gamma$  and  
parameter  $\alpha$  by  
\begin{eqnarray}   \label{a,b,c} 
\begin{array}{c}   \displaystyle   
\Gamma  \,=\,   
\frac{ \bigl| \mbox{$\bm{p}$}_{B'}^{} \bigr| }{4\pi\, m_{B}^{}} 
\bigl( E_{B'}^{}+m_{B'}^{} \bigr) 
\left( |s|^2 + |p|^2 \right)   \,\,,   
\hspace{3em}  
\alpha  \,=\,  \frac{2\,{\rm Re}(s^*p)}{|s|^2 + |p|^2}   \,\,,   
\end{array}       
\end{eqnarray}       
where  $\,s={\cal A}^{(S)}\,$  and  
$\,p=\bigl| \mbox{$\bm{p}$}_{B'}^{} \bigr| \, {\cal A}^{(P)}.\,$   
To express our results, we also adopt the notation~\cite{jenkins2}      
\begin{eqnarray}        
{a}^{(S,P)}_{BB'\phi}  \,\equiv\,  
\sqrt 2\, f^{}\, {\cal A}^{(S,P)}_{BB'\phi}   \,\,.
\label{conamp}   
\end{eqnarray}    

\begin{figure}[ht]         
\includegraphics{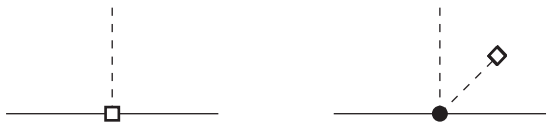}   \vspace{-3ex}
\caption{\label{Swave}%
Leading-order diagrams for chromomagnetic-penguin contributions to 
$S$-wave hyperon nonleptonic decays. 
In all figures, a solid (dashed) line denotes a baryon 
(meson) field, and a solid dot (hollow square) represents 
a strong (weak) vertex.
}
\end{figure}

\begin{figure}[ht]         
\includegraphics{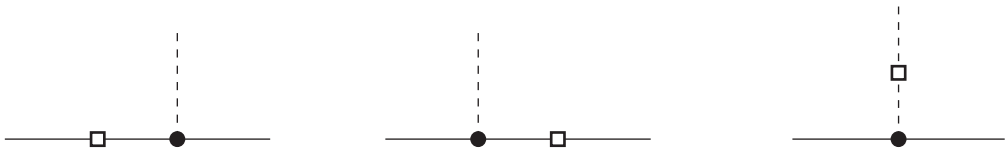}   \vspace{-3ex}
\caption{\label{Pwave}%
Leading-order diagrams for chromomagnetic-penguin contributions to 
$P$-wave hyperon nonleptonic decays.} 
\end{figure}

From the Lagrangians given above, one can derive the $S$-~and $P$-wave 
amplitudes at leading order, represented by the diagrams 
in Figs.~\ref{Swave} and~\ref{Pwave}, respectively.    
For the $S$-wave, the first diagram is directly obtained from a weak 
vertex provided by Eq.~(\ref{weakcl}), whereas the other diagram involves 
a $\bar K$-vacuum tadpole from Eq.~(\ref{weakcl}) and a strong  
$\,B$$\to$$B'\bar K\pi\,$ vertex,  which consists of contributions from both  
${\cal L}_{\rm s}^{(1)}$  and  ${\cal L}_{\rm s}^{(2)}$. 
It is worth mentioning here that the  $\beta_\varphi^{}$  and  
$\tilde\beta_\varphi^{}$  terms in  ${\cal L}_{\rm w}^{}$ 
do not contribute to $\,\bar{K}\to\pi\pi\,$  decay, as the corresponding 
direct and tadpole diagrams  cancel exactly~\cite{weinbergmod}.
For  $\,\Lambda\to p\pi^-\,$  and  $\,\Xi^-\to\Lambda\pi^-,\,$  
the resulting amplitudes are then\footnote{The contribution from  
${\cal L}_{\rm s}^{(1)}$  to the tadpole amplitude contains the factor  
$\,v\cdot p_\phi^{}=m_B^{}-m_{B'}^{}={\cal O}(m_s^{}).\,$  
As a result, the  ${\cal L}_{\rm s}^{(1)}$  and  ${\cal L}_{\rm s}^{(2)}$  
contributions to the amplitude both have the same $m_s^{}$ order, 
${\cal O}(m_s^0)$, as the $\beta_\varphi^{-}$ terms indicate.}       
%
\begin{eqnarray}   \label{stree}     
\begin{array}{c}   \displaystyle  
{a}^{(S)}_{\Lambda p\pi^-}  \,=\,  
\mbox{$\frac{1}{\sqrt{6}}$} \left( \beta_D^{-}+3 \beta_F^{-} \right)   
\,+\, 
\mbox{$\frac{3}{\sqrt{6}}$}\, \beta_\varphi^{-}\, 
\frac{m_\Lambda^{}-m_N^{}}{m_s^{}-\hat m}   \,\,,   
\vspace{2ex} \\   \displaystyle    
{a}^{(S)}_{\Xi^-\Lambda\pi^-}  \,=\,  
\mbox{$\frac{1}{\sqrt{6}}$} \left( \beta_D^{-}-3 \beta_F^{-} \right)    
\,-\, 
\mbox{$\frac{3}{\sqrt{6}}$}\, \beta_\varphi^{-}\, 
\frac{m_\Xi^{}-m_\Lambda^{}}{m_s^{}-\hat m}   \,\,,   
\end{array}    
\end{eqnarray}    
where  $\,\beta_i^{-}\equiv\beta_i^{}-\tilde{\beta}_i^{}\,$  and  
we have used the relations  
\begin{eqnarray}   \label{masses}
\begin{array}{c}   \displaystyle      
m_\Lambda^{}-m_N^{}  \,=\,  
-\mbox{$\frac{2}{3}$} \bigl( b_D^{}+3 b_F^{} \bigr) (m_s^{}-\hat m)   \,\,,     
\hspace{3em}  
m_\Xi^{}-m_\Lambda^{}  \,=\,  
\mbox{$\frac{2}{3}$} \bigl( b_D^{}-3 b_F^{} \bigr) (m_s^{}-\hat m)   \,\,,     
\vspace{1ex} \\   \displaystyle    
m_K^2  \,=\,  B_0^{}\, (m_s^{}+\hat m)   \,\,,   
\end{array}      
\end{eqnarray}      
derived from Eq.~(\ref{Ls2}).  
For the $P$-wave, the amplitude arises from two baryon-pole 
diagrams, each involving a weak vertex from Eq.~(\ref{weakcl}) and 
a strong vertex from Eq.~(\ref{Ls1}), and a kaon-pole diagram involving
a strong vertex from Eq.~(\ref{Ls1}) followed by a $\bar K$-$\pi$ 
vertex from Eq.~(\ref{weakcl}).   
Thus we find    
\begin{eqnarray}   \label{ptree}      
\begin{array}{c}   \displaystyle  
{a}^{(P)}_{\Lambda p\pi^-}     \,=\,   \displaystyle      
\frac{ (D+F) \left( \beta_D^{+}+3\beta_F^{+} \right) }{
      \sqrt{6}\,\, \bigl( m_\Lambda^{}-m_N^{} \bigr) }   
+ \frac{ 2 D \left( \beta_D^{+}-\beta_F^{+} \right) }{
        \sqrt{6}\,\, \bigl( m_\Sigma^{}-m_N^{} \bigr) }   
\,+\,  
\frac{(D+3F)\, \beta_\varphi^{+}}{\sqrt6\, \bigl(m_s^{}-\hat m\bigr) }   \,\,, 
\vspace{2ex} \\   \displaystyle    
{a}^{(P)}_{\Xi^-\Lambda\pi^-}  \,=\,   \displaystyle   
\frac{ (-D+F) \left( \beta_D^{+}-3\beta_F^{+} \right) }{  
      \sqrt{6}\,\, \bigl( m_\Xi^{}-m_\Lambda^{} \bigr) }   
- \frac{ 2 D \left( \beta_D^{+}+\beta_F^{+} \right) }{  
        \sqrt{6}\,\, \bigl( m_\Xi^{}-m_\Sigma^{} \bigr) }   
\,+\,  
\frac{(D-3F)\, \beta_\varphi^{+}}{\sqrt6\, \bigl(m_s^{}-\hat m\bigr) }   \,\,,  
\end{array} 
\end{eqnarray}    
%
where  $\,\beta_i^{+}\equiv\beta_i^{}+\tilde{\beta}_i^{}\,$  and  
we have used  $\,m_K^2-m_\pi^2=B_0^{}\, (m_s^{}-\hat m).\,$  
We note that the baryon and meson masses in all the amplitudes above 
are isospin-averaged ones.


\section{Numerical results and discussion\label{results}}
    
In order to estimate the weak phases in $A_{\Lambda,\Xi}^{}$, we first 
need to determine the parameters  $\beta_i^{}$  and~$\tilde\beta_i^{}$  
in terms of the underlying coefficients  $C_g^{}$  and~$\tilde{C}_g^{},$  
respectively.     
From the effective Hamiltonian in Eq.~(\ref{H_sdg}) and the chiral 
Lagrangian in Eq.~(\ref{weakcl}), we can derive the one-particle matrix 
elements 
\begin{eqnarray}   \label{<B'|H|B>}   
\begin{array}{c}   \displaystyle   
\bigl\langle n \bigr| {\cal H}_{\rm w}^{} \bigl| \Lambda \bigr\rangle  
\,=\,  
\frac{\beta_D^{}+3\beta_F^{}+\tilde\beta_D^{}+3\tilde\beta_F^{}}{\sqrt 6}\, 
\bar u_n^{} u_\Lambda^{}  \,\,,  
\vspace{2ex} \\   \displaystyle    
\bigl\langle \Lambda \bigr| {\cal H}_{\rm w}^{} \bigl| \Xi^0 \bigr\rangle  
\,=\,  
\frac{\beta_D^{}-3\beta_F^{}+\tilde\beta_D^{}-3\tilde\beta_F^{}}{\sqrt 6}\,  
\bar u_\Lambda^{} u_\Xi^{}   \,\,,  
\vspace{2ex} \\   \displaystyle    
\bigl\langle \pi^-\bigr| {\cal H}_{\rm w}^{} \bigl| K^-\bigr\rangle  
\,=\,  
\bigl( \beta_\varphi^{}+\tilde\beta_\varphi^{} \bigr) \, B_0^{}   \,\,.
\end{array}      
\end{eqnarray}      
Since there is presently no reliable way to calculate these matrix 
elements from first principles, we will employ the MIT bag model to 
estimate them, following earlier work~\cite{NP'}.  
Using the results given in Appendix~\ref{bagmodel}, and setting 
$\,\tilde{C}_g^{}=0,\,$  we find 
\begin{eqnarray}   \label{bdbf}  
\begin{array}{c}   \displaystyle   
\beta_D^{}  \,=\,  -\mbox{$\frac{3}{7}$}\, \beta_F^{}  \,=\,  
\frac{2\, I_M^{}N^4}{\pi\, R^2}\, C_g^{}   \,\,,  
\hspace{3em}  
\beta_\varphi^{}  \,=\,  
\frac{-8\sqrt2\, I_M^{}N^4\, m_K^{}}{\pi B_0^{}\, R^2}\, C_g^{}   \,\,,   
\end{array}      
\end{eqnarray}      
where  $N$, $R$, and $I_M^{}$  are bag parameters whose values are 
given in the appendix.   
By setting  $\,C_g^{}=0\,$  instead, one finds similar relations 
between  $\tilde\beta_i^{}$  and  $\tilde{C}_g^{}$.  
It follows that numerically 
\begin{eqnarray}   \label{beta_i}  
\begin{array}{c}   \displaystyle   
\stackrel{\mbox{\tiny$\scriptscriptstyle\!(\sim)$}}{\beta_D^{}}  \,=\,  
-\mbox{$\frac{3}{7}$}   
\stackrel{\mbox{\tiny$\scriptscriptstyle\!(\sim)$}}{\beta_F}{}^{}  
\,=\,  
1.10\times10^{-3}\,\, 
\stackrel{\mbox{\tiny$\scriptscriptstyle\!(\sim)$}}{C_g^{}}\,\,  
{\rm GeV}^2   \,\,,  
\hspace{3em}  
\stackrel{\mbox{\tiny$\scriptscriptstyle\!(\sim)$}}{\beta_\varphi^{}}  \,=\, 
-3.49\times10^{-3}\,\, 
\stackrel{\mbox{\tiny$\scriptscriptstyle\!(\sim)$}}{C_g^{}}\,\, 
{\rm GeV}^2   \,\,,  
\end{array}      
\end{eqnarray}      
where we have used  $\,B_0^{}=m_K^2/(m_s^{}+\hat m),\,$  with  
$\,m_K^{}=495.7\,\rm MeV\,$  and~\cite{cpreview}  $\,m_s^{}+\hat m=121\,{\rm MeV}.\,$  
We note that  $C_g^{}$  and  $\tilde C_g^{}$  here are the Wilson 
coefficients at the low scale  $\,\mu={\cal O}(1\,\rm GeV)\,$  and hence 
already contain the QCD running from the new-physics scales. 
We also note that the bag-model numbers in Eq.~(\ref{beta_i}) are comparable in 
magnitude to the natural values of the parameters as obtained from naive 
dimensional analysis~\cite{nda}, e.g.,   
\begin{eqnarray}   \label{beta_i^nda}   
\beta_{D,F}^{\rm NDA}  \,=\,  
\frac{C_g^{}\, g_{\rm s}^{}}{16\pi^2}\, \frac{\Lambda^2}{4\pi}   
\,\sim\,  0.0024\,C_g^{}\,\, {\rm GeV}^2   \,\,,  
\hspace{3em}  
\beta_\varphi^{\rm NDA}  \,=\,  {C_g^{}\, g_{\rm s}^{}\over 16\pi^2}\, 
\frac{\Lambda^3}{4\pi B_0^{}}   
\,\sim\,  0.0014\,C_g^{}\,\, {\rm GeV}^2   \,\,,  
\end{eqnarray}      
where  $\,\Lambda=4\pi f\,$  is the chiral-symmetry breaking 
scale,  with  $\,f=f_\pi^{}\simeq 92.4\,\rm MeV,\,$  
and  we have chosen  $\,g_{\rm s}^{}=\sqrt{4\pi}.\,$  
The differences between the two sets of numbers provide an indication 
of the level of uncertainty in estimating the matrix elements.
This will be taken into account in our results below.

Next, we adopt the usual prescription for obtaining a weak 
phase~\cite{dhp,tv,NP'}, namely dividing the imaginary part of 
the theoretical amplitude by the real part of the amplitude extracted 
from experiment under the assumption of no $CP$ violation.  
For the real part of amplitudes, we employ the experimental values 
obtained in Ref.~\cite{at}.
They are, in units of  $G_{\rm F}^{}m_{\pi^+}^2$,  
\begin{eqnarray}   \label{spx}
\begin{array}{c}   \displaystyle   
s_{\Lambda\to p\pi^-}^{}= 1.42\pm0.01   \,\,,   \hspace{3em}  
s_{\Xi^-\to\Lambda\pi^-}^{}= -1.98 \pm 0.01   \,\,,   
\vspace{2ex} \\   \displaystyle    
p_{\Lambda\to p\pi^-}^{}= 0.52\pm 0.01   \,\,,   \hspace{3em}  
p_{\Xi^-\to\Lambda\pi^-}^{}= 0.48\pm 0.02   \,\,.  
\end{array}      
\end{eqnarray}      
The imaginary part of the amplitudes are calculated from  
Eqs.~(\ref{stree}) and~(\ref{ptree}), combined with  Eq.~(\ref{beta_i}).     
The other hadron masses that we employ are  $\,m_N^{}=938.9,\,$  
$\,m_\Lambda^{}=1115.7,\,$  $\,m_\Sigma^{}=1193.2,\,$  
$\,m_\Xi^{}=1318.1,\,$  and  $\,m_\pi^{}=137.3,\,$  all in units of MeV.   
In Table~\ref{ratios}, we have collected the results, divided 
by the central values of Eq.~(\ref{spx}), in terms of~$C_g^{}$  
and~$\tilde{C}_g^{}$.  
We find that in each of the amplitudes the  $\beta_\varphi^\pm$  terms  
are numerically larger than the  $\beta_{D,F}^\pm$  terms,  
by up to a factor of four, and both contribute with the same sign.

\begin{table}[ht]   
\caption{\label{ratios}%
Ratios of theoretical amplitude arising from chromomagnetic operators 
to experimental amplitude, for $S$- and $P$-wave transitions.  
}   
\centering   \footnotesize
\vskip 0.5\baselineskip
\begin{tabular}{@{\hspace{1em}}c@{\hspace{1em}}|@{\hspace{1em}}c
@{\hspace{3em}}c@{\hspace{1em}}}   
\hline \hline 
Decay mode $\vphantom{\Biggl|_|^|}$  &  
$\displaystyle \frac{{\rm Im}\,s}{s_{\rm expt}^{}}$ (GeV) &    
$\hspace{1ex}\displaystyle 
\frac{{\rm Im}\,p}{p_{\rm expt}^{}}$ (GeV) 
\\ \hline 
$\Lambda\to p\pi^-  \vphantom{\Big|_|^|}$  &  
$-2.2\times10^5\,{\rm Im} \bigl(C_g^{}-\tilde{C}_g^{}\bigr)$  &  
$-2.6\times10^5\,{\rm Im} \bigl(C_g^{}+\tilde{C}_g^{}\bigr)$  \\ 
$\Xi^-\to\Lambda\pi^-  \vphantom{\Big|_|^|}$  &  
$-1.9\times10^5\,{\rm Im} \bigl(C_g^{}-\tilde{C}_g^{}\bigr)$  &  
$+1.1\times10^5\,{\rm Im} \bigl(C_g^{}+\tilde{C}_g^{}\bigr)$  
\\ \hline \hline  
\end{tabular}    
\end{table}   

Since the ratios in Table~\ref{ratios} follow  from the leading-order 
amplitudes in~$\chi$PT, the uncertainty of our prediction will come partly 
from our lack of knowledge about the higher-order terms which are 
presently incalculable.    
Various studies of hyperon processes in the context of  $\chi$PT  
show that the leading nonanalytic contributions to amplitudes can be 
comparable to the lowest-order terms~\cite{bsw,JenMan,jenkins2,at,nlhd}. 
We expect that a similar situation occurs here, and consequently we also 
expect the uncertainty due to the higher-order contributions to be  
comparable to our leading-order estimate.   
To reflect this, as well as the uncertainty in estimating the matrix 
elements above, we assign an error of 200$\%$ to each of these ratios.  
In Table~\ref{phases}, we have listed the ratios as the weak phases,\footnote{
We remark here that the central values of the numerical factors 
in front of ${\rm Im}\, C_g^\pm$  for  $\phi_{S,P}^{\Lambda}$ in  
Table~\ref{phases} are larger, by roughly a factor of two, than the corresponding 
numbers obtained in Ref.~\cite{NP'}, which considers contributions from  
a generic  supersymmetric model.   
The disagreement may be due mainly to a factor-of-two difference between 
the  matrix elements in Eqs.~(\ref{<B'|sdG|B>})  and~(\ref{<pi|sdG|K>})  and 
those employed in  Ref.~\cite{NP'}.        
} 
along with their uncertainties, in terms of  
$\,C_g^\pm\equiv C_g^{}\pm\tilde{C}_g^{}.\,$    
Accordingly,  $C_g^+$  and  $C_g^-$  correspond to  parity-even 
and parity-odd  transitions, respectively.       

\begin{table}[ht]   
\caption{\label{phases}%
Weak phases generated by chromomagnetic operators   
and strong-phase differences,  $\delta_S^{}-\delta_P^{}$.  
}   
\centering   \footnotesize
\vskip 0.5\baselineskip
\begin{tabular}{c||@{\hspace{1em}}c@{\hspace{2em}}c
@{\hspace{1em}}|@{\hspace{1em}}c}   
\hline \hline 
Decay mode $\vphantom{\Biggl|_o^o}$  &  
$10^{-5}\, \phi_S^{}$ (GeV) &  $10^{-5}\,\phi_P^{}$ (GeV) &  
$\hphantom{-}\delta_S^{}-\delta_P^{}\hphantom{-}$ 
\\ \hline 
$\Lambda\to p\pi^-  \vphantom{\Big|_|^|}$  &  
$(-2.2\pm 4.4)\,{\rm Im}\, C_g^{-}$  &  
$(-2.6\pm 5.2)\,{\rm Im}\, C_g^{+}$  &   
$\hphantom{-}7^\circ\pm2^\circ$ 
\\     
$\Xi^-\to\Lambda\pi^-  \vphantom{\Big|_|^|}$  &  
$(-1.9\pm 3.8)\,{\rm Im}\, C_g^{-}$  &  
$\hphantom{-}( 1.1\pm 2.2)\,{\rm Im}\, C_g^{+}$  & 
$-2^\circ\pm 6^\circ$   
\\
\hline \hline  
\end{tabular}    
\end{table}

In Table~\ref{phases}, we have also included the strong-phase differences. 
The number for $\,\Lambda\to p\pi^-\,$  results from the measured 
phases quoted in Sec.~\ref{intro}.  
For $\,\Xi^-\to\Lambda\pi^-,\,$  while awaiting a definitive measurement 
by HyperCP, we have adopted the range 
$\,-7.8^\circ<\delta_{S}^{\Xi}-\delta_{P}^{\Xi}<+3.9^\circ\,$ 
estimated in Ref.~\cite{ttv}.   
This range is compatible with the experimental values known to date, including 
the preliminary measurement by HyperCP mentioned in Sec.~\ref{intro}.

From the results in Table~\ref{phases}, it follows that the contributions 
of the CMO are  
\begin{eqnarray}   
\begin{array}{c}   \displaystyle   
10^{-4}\, \bigl(A_\Lambda^{}\bigr)_g  \,=\,  
(3.5\pm 7.0)\, {\rm Im}\,C_g^{-} + (-4.2\pm 8.3)\, {\rm Im}\,C_g^{+}   \;,
\vspace{2ex} \\   \displaystyle    
10^{-4}\, \bigl(A_\Xi^{}\bigr)_g  \,=\,  
(-2.0\pm 6.0)\, {\rm Im}\,C_g^{-} + (-1.2\pm 3.4)\, {\rm Im}\,C_g^{+}   \;.  
\end{array}      
\end{eqnarray}      
where the numbers on the right-hand sides are all in units of GeV.   
This indicates that  $(A_\Xi^{})_g^{}$  is not negligible compared to 
$(A_\Lambda^{})_g^{}$  and hence should be included in evaluating  
$A_{\Xi\Lambda}^{}$.   
Summing the two asymmetries then yields     
\begin{eqnarray}   \label{A_XL}   
10^{-4}\, \bigl(A_{\Xi\Lambda}^{}\bigr)_g  \,=\,  
(2\pm 13)\, {\rm Im}\,C_g^{-} + (-5\pm 12)\, {\rm Im}\,C_g^{+}   \;,
\end{eqnarray}      
the right-hand side being again in GeV.   
The errors we quote here are obviously not Gaussian, and simply indicate 
the ranges resulting from our calculation of the phases.

Since the CMO also contribute to the parameters 
$\epsilon'$  and  $\epsilon$  in the kaon sector,  it is possible to 
obtain a bound on their contribution  to  $A_{\Xi\Lambda}^{}$  
using the measured values of  $\epsilon'$  and  $\epsilon$.  
The contribution to  $\epsilon'$  can be written as~\cite{buras,NP'}
\begin{eqnarray}   \label{e'}
\left(\frac{\epsilon'}{\epsilon}\right)_{\!\!g}  \,=\,  
\Bigl( 5.2\times10^5\,{\rm GeV} \Bigr) \, B_G^{}\,\, {\rm Im} C_g^-   \,\,,  
\end{eqnarray}      
where  $B_G^{}$  parameterizes the hadronic uncertainty, and 
$\,m_s^{}+\hat m=121\,{\rm MeV}\,$ has been used~\cite{cpreview}. 
The contribution to  $\epsilon$  occurs through long-distance effects, 
and the simplest ones arise from $\pi^0$,  $\eta$,  and  $\eta'$   
poles~\cite{weinbergmod}, yielding  
\begin{eqnarray}   \label{e}   
( \epsilon )_{g}^{}  \,=\,  
-\Bigl( 2.3\times10^5\,{\rm GeV} \Bigr) \, \kappa\,\, {\rm Im} C_g^+   \,\,,
\end{eqnarray}      
where  $\kappa$  quantifies the contributions of the different poles.  
Hence Eq.~(\ref{A_XL}) can be rewritten as  
\begin{eqnarray}   \label{A_XL'}
\bigl(A_{\Xi\Lambda}^{}\bigr)_g  \,=\,  
\frac{0.04\pm 0.25}{B_G^{}} \left(\frac{\epsilon'}{\epsilon}\right)_{\!\!g} 
\,+\, \frac{0.22\pm 52}{\kappa}\, (\epsilon)_{g}^{}   \;.
\hspace*{2em} 
\end{eqnarray}      
To estimate the range of  $\bigl(A_{\Xi\Lambda}^{}\bigr){}_g$  allowed 
by the experimental values  
$\,|\epsilon|=(22.80\pm 0.13)\times10^{-4}\,$   and  
$\,{\rm Re}(\epsilon'/\epsilon)=(16.6\pm1.6)\times10^{-4}\,$~\cite{cpreview,pdb},   
we require  
\begin{eqnarray}   \label{ee'bound}
\left(\frac{\epsilon'}{\epsilon}\right)_{\!\!g} \,<\,  19\times10^{-4}   \,\,, 
\hspace{3em}  
|\epsilon|_g^{}  \,<\,  23\times10^{-4}  \,\,.   
\end{eqnarray}      
Consequently, adopting  $\,0.5<B_G^{}<2\,$  and  $\,0.2<|\kappa|<1,\,$  after   
Ref.~\cite{NP'}, we find the bound  
\begin{eqnarray}   \label{A_XLg} 
\bigl|A_{\Xi\Lambda}^{}\bigr|_g  \,<\,  97\times10^{-4}   \,\,.  
\end{eqnarray}      
The upper limit of this range is allowed by the published data~\cite{pdb},  
but is disfavored by the preliminary result of HyperCP~\cite{zyla} 
quoted in Sec.~\ref{intro}, exceeding it by several sigmas.  
Since the number in Eq.~(\ref{A_XLg}) is dominated by the  $(\epsilon)_g^{}$ 
bound, we can then conclude that the available preliminary measurement by  
HyperCP already probes the parity-even contributions better than  $\epsilon$  
does.

Now, it is possible that HyperCP will in the end observe no $CP$-violation 
in  $A_{\Xi\Lambda}^{}$.  
In that event, the data can be used to estimate the bounds on both  ${\rm Im}C_g^\pm$.    
To explore this possibility, we assume that HyperCP will be able to reach 
the expected sensitivity of $\,2\times10^{-4}\,$~\cite{zyla}, and 
so we take this number as the upper limit for  $A_{\Xi\Lambda}^{}$.  
Moreover, since our result in Eq.~(\ref{A_XL}) has large uncertainties, 
for illustrative purposes we use its central value in what follows.    
Barring significant cancellations between the  ${\rm Im}C_g^\pm$  terms,   
we can consider three possible cases,  
(i)~${\rm Im}C_g^+=0$  and~${\rm Im}C_g^-\neq0,\,$ 
(ii)~${\rm Im}C_g^+\neq0$  and~${\rm Im}C_g^-=0,\,$  and  
(iii)~${\rm Im}C_g^+\sim -{\rm Im}C_g^-\neq0.\,$   
Consequently, requiring  $\,|A_{\Xi\Lambda}^{}|_g^{} < 2\times10^{-4},\,$  
we obtain for these cases  
\begin{eqnarray}   
\begin{array}{c}   \displaystyle   
{\rm (i)}~\,\bigl|{\rm Im}C_g^-\bigr| \,\lesssim\, 1\times10^{-8}\,{\rm GeV}^{-1}  \,\,,  
\hspace{3em}  
{\rm (ii)}~\,\bigl|{\rm Im}C_g^+\bigr| \,\lesssim\, 4\times10^{-9}\,{\rm GeV}^{-1}  \,\,,   
\vspace{2ex} \\   \displaystyle    
{\rm (iii)}~\,\bigl|{\rm Im}C_g^+\bigr|  \,\sim\,   
\bigl|{\rm Im}C_g^-\bigr|  \,\lesssim\,  3\times10^{-9}\,{\rm GeV}^{-1}   \,\,.   
\end{array}      
\end{eqnarray}      
For comparison, the requirements in Eq.~(\ref{ee'bound}) from  
$\epsilon'$  and  $\epsilon$  measurements imply   
\begin{eqnarray}   
\bigl|{\rm Im}C_g^-\bigr|  \,<\,  7.4 \times10^{-9}\,{\rm GeV}^{-1}   \,\,,  
\hspace{3em}  
\bigl|{\rm Im}C_g^+\bigr|  \,<\,  5.0\times10^{-8}\,{\rm GeV}^{-1}   \,\,.   
\end{eqnarray}      
Similar or stronger bounds on  ${\rm Im}C_g^+$  may also be obtainable from  
$\,K\to3\pi,\,\pi\ell^+\ell^-,\,\pi\gamma\gamma\,$  decays~\cite{evenparity}.
Therefore, even if it turns out that HyperCP eventually does not discover $CP$ 
violation in hyperon decays,  its data  can be expected to provide stringent 
constraints on the coefficients of the CMO in new-physics models, at a level 
that is comparable to or better than the bounds coming from the kaon sector.

\section{Conclusion\label{conclusion}}
   
We have evaluated the $CP$-violating asymmetries  $A_\Lambda^{}$  and  
$A_\Xi^{}$  induced by the chromomagnetic-penguin operators, 
whose coefficients can be enhanced in some scenarios of new physics.  
Including recent information on the strong phases in  $\,\Xi\to\Lambda\pi\,$  
and adopting a model-independent approach, we have shown that  
$\bigl(A_\Xi^{}\bigr){}_g$,  which was neglected in earlier studies,  
can be comparable to  $\bigl(A_\Lambda^{}\bigr){}_g$.  
We have found that the upper limit of the sum of these asymmetries, 
$\bigl(A_{\Xi\Lambda}^{}\bigr){}_g$, as allowed by  $\epsilon$  and  
$\epsilon'$  data is already disfavored by the preliminary measurement of 
$A_{\Xi\Lambda}^{}$  by HyperCP.  
It follows that this preliminary data already imposes a constraint on 
the parity-even contributions of the operators that is stronger 
than the bound obtained from  $\epsilon$  in kaon mixing.    
We have explored how well the upcoming results from HyperCP may bound 
the coefficients of the operators in the event of null results.   
In that case, the data will likely yield significant constraints that are 
comparable to or better than those provided by kaon measurements.

\begin{acknowledgments}    
I would like to thank G. Valencia for helpful comments and suggestions, 
and E.C. Dukes for some experimental information.  
This work was supported in part by the Lightner-Sams Foundation.   
\end{acknowledgments}   
   
\appendix  
 
\section{Chiral realization of 
\mbox{\boldmath$\,(\bar 3_{\rm L}^{},3_{\rm R}^{})\,$}  and  
\mbox{\boldmath$\,(3_{\rm L}^{},\bar 3_{\rm R}^{})\,$}  
operators\label{qbarq}} 
  
The form of the weak Lagrangian in Eq.~(\ref{weakcl}) can be inferred 
from the lowest-order chiral realization of the operators      
$\,\bar d(1+\gamma_5^{})s\,$  and  $\,\bar d(1-\gamma_5^{})s.\,$  
The reason is that these densities, like the operators  $Q_g^{}$  
and  $\tilde Q_g^{}$  in Eq.~(\ref{H_sdg}), transform as  
$\,\bigl(\bar 3_{\rm L}^{},3_{\rm R}^{}\bigr)\,$  and  
$\,\bigl(3_{\rm L}^{},\bar 3_{\rm R}^{}\bigr),\,$  respectively,  
under  $\,\rm SU(3)_{L}^{}$$\times$$\rm SU(3)_{R}^{}\,$  rotations.   
Using  ${\cal L}_s^{(2)}$  in Eq.~(\ref{Ls2}), one can derive   
the correspondences~\cite{tv}  
\begin{eqnarray}   
-\bar d_{\rm L}^{} s_{\rm R}^{}  &\Longleftrightarrow&
b_D^{}\, 
\bigl( \xi^\dagger B_v^{} \bar B_v^{} \xi^\dagger 
      + \xi^\dagger \bar B_v^{} B_v^{} \xi^\dagger \bigr) _{32}^{}
\,+\,  
b_F^{}\, 
\bigl( \xi^\dagger B_v^{} \bar B_v^{} \xi^\dagger   
      - \xi^\dagger \bar B_v^{} B_v^{} \xi^\dagger \bigr)_{32}^{}  
\,+\,  
b_0^{}\, \Sigma_{32}^\dagger\, 
\bigl\langle \bar B_v^{} B_v^{} \bigr\rangle   
\nonumber \\ &&  
+\,\,  
\mbox{$\frac{1}{2}$} f^2 B_0^{}\, \Sigma_{32}^\dagger   \,\,,   
\end{eqnarray}      
\begin{eqnarray}   
-\bar d_{\rm R}^{} s_{\rm L}^{}  &\Longleftrightarrow&
b_D^{} \left( \xi B_v^{} \bar B_v^{} \xi 
             + \xi \bar B_v^{} B_v^{} \xi \right) _{32}^{}
\,+\,  
b_F^{} \left( \xi B_v^{} \bar B_v^{} \xi   
             - \xi \bar B_v^{} B_v^{} \xi \right)_{32}^{}  
\,+\,  
b_0^{}\, \Sigma_{32}^{}\, 
\bigl\langle \bar B_v^{} B_v^{} \bigr\rangle   
\nonumber \\ &&  
+\,\,  
\mbox{$\frac{1}{2}$} f^2 B_0^{}\, \Sigma_{32}^{}   \,\,, 
\end{eqnarray}      
where  $\,q_{\rm L}^{}=\frac{1}{2}(1-\gamma_5)q\,$ 
and  $\,q_{\rm R}^{}=\frac{1}{2}(1+\gamma_5)q.\,$   
The form in Eq.~(\ref{weakcl}) then follows.

It is worth noting here that each of the $S$- and $P$-wave amplitudes in 
Eqs.~(\ref{stree}) and~(\ref{ptree}) vanishes if we set    
\begin{eqnarray}   
\beta_{D,F}^{} \,=\,  {c}\, b_{D,F}^{}   \,\,,  \hspace{2em} 
\beta_\varphi^{} \,=\,  \frac{{c}}{2}  \,\,,  \hspace{2em}   
\tilde\beta_{D,F}^{} \,=\,  \tilde{c}\, b_{D,F}^{}  \,\,,  \hspace{2em} 
\tilde\beta_\varphi^{} \,=\,  \frac{\tilde{c}}{2}  \,\,,   
\end{eqnarray}      
with ${c}$ and  $\tilde{c}$  being constants, 
and also use the relations in Eq.~(\ref{masses}) as well as  
\begin{eqnarray}   \label{masses'}
\begin{array}{c}   \displaystyle      
m_\Sigma^{}-m_N^{}  \,=\,  
2 \bigl( b_D^{}-b_F^{} \bigr) (m_s^{}-\hat m)   \,\,,     
\hspace{3em}  
m_\Xi^{}-m_\Sigma^{}  \,=\,  
-2 \bigl( b_D^{}+b_F^{} \bigr) (m_s^{}-\hat m)   \,\,, 
\end{array}      
\end{eqnarray}      
derived from Eq.~(\ref{Ls2}).  
This satisfies the requirement implied by the Feinberg-Kabir-Weinberg 
theorem~\cite{fkw} that the operators  $\,\bar d(1\pm\gamma_5^{})s\,$  
cannot contribute to physical amplitudes~\cite{DonGH1}, and thus serves 
as a check for the formulas in Eqs.~(\ref{stree}) and~(\ref{ptree}).

\section{Bag-model parameters\label{bagmodel}}  
  
Here we provide the estimate in the MIT bag model of 
the matrix elements of the chromomagnetic operators contained in 
Eq.~(\ref{<B'|H|B>}).    
The relevant calculations can be found in Refs.~\cite{DonGH2,dghp}.  
We have   
\begin{eqnarray}   \label{<B'|sdG|B>}         
\begin{array}{c}   \displaystyle   
\bigl\langle n \bigl| g_{\rm s}^{}\, \bar d \sigma^{\mu\nu} \lambda^a 
\bigl(1\pm\gamma_5^{}\bigr) s\, G_{\!\mu\nu}^{a} \bigr| \Lambda \bigr\rangle 
\,=\,   
\frac{-16\sqrt6\,\, g_{\rm s}^2\, N^4 I_M^{}}{R^2}\, 
\bar u_{n}^{} u_\Lambda^{}   \,\,,   
\vspace{2ex} \\   \displaystyle    
\bigl\langle \Lambda \bigl| g_{\rm s}^{}\, \bar d\sigma^{\mu\nu} \lambda^a 
\bigl(1\pm\gamma_5^{}\bigr) s\, G_{\!\mu\nu}^{a} \bigr| \Xi^0 \bigr\rangle 
\,=\,   
\frac{64\sqrt6\,\, g_{\rm s}^2\, N^4 I_M^{}}{3\, R^2}\,  
\bar u_\Lambda^{} u_\Xi^{}    \,\,,    
\end{array}      
\end{eqnarray}      
\begin{eqnarray}   \label{<pi|sdG|K>}   
\bigl\langle \pi^- \bigl| g_{\rm s}^{}\, \bar d \sigma^{\mu\nu} \lambda^a 
\bigl(1\pm\gamma_5^{}\bigr) s\, G_{\!\mu\nu}^{a} \bigr| K^- \bigr\rangle 
\,=\,  \frac{-64\, g_{\rm s}^2\, N^4 I_M^{}}{R^2}\, \sqrt{2 m_K^2}   \,,         
\end{eqnarray}      
where  $\,\lambda^a=2t^a,\,$  only the parity-conserving part of 
the $\,s\to d g\,$  operators contributes, and 
$R$, $N$, and  $I_M^{}$  are bag-model parameters~\cite{dghp}.  
Numerically, we choose  $\,g_{\rm s}^2=4\pi,\,$   
corresponding to  $\,\alpha_{\rm s}^{}=1,\,$  and adopt  
$\,R=5.0\,{\rm GeV}^{-1}\,$  for the baryons  and  
$\,R=3.3\,{\rm GeV}^{-1}\,$  for the mesons~\cite{dghp}. 
Since the weak parameters  $\beta_i^{}$ and  $\tilde\beta_i^{}$  
belong to a Lagrangian which respects SU(3) symmetry  [${\cal L}_{\rm w}$  
in Eq.~(\ref{weakcl})],  in writing  Eqs.~(\ref{<B'|sdG|B>})  
and~(\ref{<pi|sdG|K>})  we have employed SU(3)-symmetric kinematics.   
Accordingly, we take  $\,m_u^{}=m_d^{}=m_s^{}=0\,$  and  use 
the formulas given in Ref.~\cite{dghp} to obtain  $\,N=2.27\,$  and  
$\,I_M^{}=1.63\times10^{-3}\,$   for both the baryons and mesons.    
Finally, we note that Eq.~(\ref{<pi|sdG|K>}),  together with the relation 
$\,\langle\pi^-|Q_g|K^-\rangle=-\sqrt2\, \langle\pi^0|Q_g|\bar K^0\rangle,\,$  
leads to the matrix element~\cite{weinbergmod,DonHH}
$\,A_{\bar K\pi}^{}\equiv
\bigl\langle \pi^0 \bigl| \bar d \sigma^{\mu\nu} \lambda^a 
(1+\gamma_5^{}) s\, G_{\!\mu\nu}^{a} \bigr| \bar K{}^0 \bigr\rangle 
=+64\, g_{\rm s}^{}\, N^4 I_M^{}\, m_K^{}/R^2 \simeq 0.4\,\rm GeV^3 .\,$

\end{document}